\input harvmac

\def \const {{\rm const}}

\def \e {{\rm e}} 
\def \ov {\over}

\def \ep {\epsilon}
\def \k {\kappa}

\def \L {{\cal L}}

\def \J {{\cal J}}

\def \ss {{\cal S}}

\def \a {\alpha}
\def \E {{\cal E}}

\def \g {\gamma}
\def \G {\Gamma}
\def \d {\delta}

\def \l {\lambda}

\def \n {\nu}

\def \s {\sigma}

\def \eps {\epsilon}

\def \t {\theta}
\def \ta {\tau}

\def \vp {\varphi}

\def \frac#1#2{{ #1 \over #2}}
\def \lr { \lref}
\def \td {\tilde}

\def \lr{\lref}

\def \rf {\refs}

\def \adss {$AdS_5 \times S^5\ $}

\def \ta { \tau}
\def \s { \sigma }

\def \vp {\varphi}

\def \vt {\theta}
\def \bx {\bar x} \def \a { \alpha}

\def \fourth {{1 \ov 4}}

\def \vr {\varrho}
  \def \td { \tilde }

\def \t {\theta}

\def \del{\partial}

\def \n {\nu }
\def \ha { { 1 \over 2}}

\def \vr  { \varrho}

\def \g {\gamma}
\def \G {\Gamma}
\def \k {\kappa}
\def \l {\lambda}
\def \L {{\cal L}}

\def \td {\tilde }

\def \tr {{\rm tr}}
\def \ha {{1 \over 2}}
\def \ep{\epsilon}

\def \ov {\over}
\def \JJ {{\cal J}} \def \Om {\Omega}

\def \w {w}

\def \om {\omega}
\def \w  {\omega}

\def \ww { {\rm w} }

\def \sql {{\sqrt{\l}}\ }

\def \ta {\tau}

\def \D {{\Delta}}
\def \E {{\cal E}}
\def \JJ {{J'}}
\def \ta {\tau}

\def \L {\Lambda}
\def \II {{\cal V}}

\def \rD {{\rm D}}

\def \X {{\rm  X}}
\def \JJ {{ \cal I}}

\def \bx {{\bar X}}
\def \Om {\Omega}
\def \ss { \sin^2 \g_0\ }
\def \E {{\cal E}}
\def \kk {{\rm k}}
\def \cc {{\cos^2 \g_0}}
\def \ss {{\sin^2 \g_0}}
\def \hw {{\alpha}}

\lr \papa {
M.~Blau, J.~Figueroa-O'Farrill, C.~Hull and G.~Papadopoulos,
``A new maximally supersymmetric background of IIB superstring theory,''
JHEP {\bf 0201}, 047 (2002)
[hep-th/0110242].
}

\lr \bmn { D.~Berenstein, J.~Maldacena and H.~Nastase,
``Strings in flat space and pp waves from N = 4 super Yang
 Mills,''
JHEP {\bf 0204}, 013 (2002)
[hep-th/0202021].
}

\lr \gkp {
S.~S.~Gubser, I.~R.~Klebanov and A.~M.~Polyakov,
``A semi-classical limit of the gauge/string correspondence,''
Nucl.\ Phys.\ B {\bf 636}, 99 (2002)
[hep-th/0204051].
}

\lref\mets{
R.~R.~Metsaev,
``Type IIB Green-Schwarz superstring in plane wave Ramond-Ramond
background,''
Nucl.\ Phys.\ B {\bf 625}, 70 (2002)
[hep-th/0112044].
R.~R.~Metsaev and A.~A.~Tseytlin,
``Exactly solvable model of superstring in plane wave
 Ramond-Ramond
background,''
Phys.\ Rev.\ D {\bf 65}, 126004 (2002)
[hep-th/0202109].
}

\lref\MT{
R.~R.~Metsaev and A.~A.~Tseytlin,
``Type IIB superstring action in \adss  background,''
Nucl.\ Phys.\ B {\bf 533}, 109 (1998)
[hep-th/9805028].
}

\lr \tse { A.~A.~Tseytlin,
``Semiclassical quantization of superstrings: \adss and
 beyond,''
Int.\ J.\ Mod.\ Phys.\ A {\bf 18}, 981 (2003)
[hep-th/0209116].
}

\lr \fts{ S.~Frolov and A.~A.~Tseytlin,
``Semiclassical quantization of rotating superstring in
\adss,''
JHEP {\bf 0206}, 007 (2002)
[hep-th/0204226].
}

\lr \mz{
J.~A.~Minahan and K.~Zarembo,
``The Bethe-ansatz for N = 4 super Yang-Mills,''
JHEP {\bf 0303}, 013 (2003)
[hep-th/0212208].
}

\lr \papd{
M.~Blau, M.~O'Loughlin, G.~Papadopoulos and A.~A.~Tseytlin,
``Solvable models of strings in homogeneous plane wave
backgrounds,''
hep-th/0304198.
}

\lref\ft{
S.~Frolov and A.~A.~Tseytlin,
``Multi-spin string solutions in \adss,''
hep-th/0304255.
}

\lr \plan{ J.~G.~Russo and A.~A.~Tseytlin,
``On solvable models of type IIB superstring in NS-NS and R-R
 plane
wave  backgrounds,''
JHEP {\bf 0204}, 021 (2002)
[hep-th/0202179].
R.~Corrado, N.~Halmagyi, K.~D.~Kennaway and N.~P.~Warner,
``Penrose limits of RG fixed points and pp-waves with background
fluxes,''
hep-th/0205314.
D.~Brecher, C.~V.~Johnson, K.~J.~Lovis and R.~C.~Myers,
``Penrose limits, deformed pp-waves and the string duals of N = 1
large N  gauge theory,''
JHEP {\bf 0210}, 008 (2002)
[hep-th/0206045].
}

\lr \bis{N.~Beisert, C.~Kristjansen, J.~Plefka, G.~W.~Semenoff
 and
M.~Staudacher,
``BMN correlators and operator mixing in N = 4 super Yang-Mills
theory,''
Nucl.\ Phys.\ B {\bf 650}, 125 (2003)
[hep-th/0208178].
}

\lr \mzn{N.~Beisert, J. Minahan,
M.~Staudacher and K. Zarembo, hep-th/0306139,  to appear.}

\Title{\vbox
{\baselineskip 10pt
{
}}}
{\vbox{\vskip -30 true pt
\medskip
\centerline {Quantizing three-spin  string  solution
  in $AdS_5 \times S^5$ }
\medskip
\vskip4pt }}
\vskip -20 true pt
\centerline{S. Frolov$^{a,}$\footnote{$^*$} {Also at Steklov
Mathematical Institute, Moscow.}
 and
A.A.~Tseytlin$^{a,b,}$\footnote{$^{**}$}
{Also at
Lebedev Physics Institute, Moscow.}
}
\smallskip\smallskip
\centerline{ $^a$ \it  Department of Physics,
 The Ohio State University,
 Columbus, OH 43210, USA}

\centerline{ $^b$ \it  Blackett Laboratory,
 Imperial College,
 London,  SW7 2BZ, U.K.}

\bigskip\bigskip
\centerline {\bf Abstract}
\baselineskip12pt
\noindent
\medskip

As was recently found in hep-th/0304255,  there exists  a simple
classical solution  describing a closed  string rotating in
$S^5$ and located at the center of $AdS_5$.
It is parametrized by the angular momentum $J$ of the
center of mass  and two equal $SO(6)$  angular momenta
$J'$ in the two  other orthogonal  rotation planes.
The corresponding dual $N=4$ SYM operators should
be scalar operators in  $SU(4)$ representations
$[0,J-J',2J']$ if $J\geq J'$, or  $[J'-J,0,J'+J]$ if $J'\geq J$.
This solution is stable if $J' \leq {3 \ov 2 } J$  and
for large $J + 2 J'$ its classical energy admits an expansion in
positive powers of
${ \lambda \ov (J + 2 J')^2}$ \ ($\sql$ is proportional to string
tension):
$E= J + 2 J' + { \lambda \ov (J + 2 J')^2} J' + ... $.
This suggests a possibility of a direct comparison with
perturbative SYM  results
for the corresponding anomalous dimensions in the sector with
${ \lambda \ov (J + 2 J')^2} \ll 1, $ by analogy with the BMN
case.  We  conjecture that all  quantum  sigma
model string corrections are then  subleading at large $J'$, so
that the classical formula for the energy  is effectively exact
to  all orders in  $\l$. It    could then be  interpolated to
 weak
coupling,  representing  a prediction for the anomalous
dimensions on the SYM side. We test this conjecture by computing
the 1-loop  superstring sigma model correction to the classical
energy.

\bigskip

\Date{06/03}

\noblackbox
\baselineskip 16pt plus 2pt minus 2pt


\newsec{Introduction}

Motivated by attempts \rf{\bmn,\gkp}
to extend AdS/CFT duality to non-BPS states
we have recently proposed \ft\ to
study the \adss string -- $N=4$ SYM
duality in a new sector parametrised by several
components of $S^5$ spin or several ``R-charges''.

 We have found
a new classical   solution
describing a circular closed string located
at the origin of $AdS_5$ space and
rotating  in $S^5$
 with two equal angular momenta
  in the two orthogonal planes:
the
rotating string moves on $S^3$ within $S^5$
just as in the case of  a two-spin flat-space  solution where
the string rotates in two orthogonal planes while always  lying
on a  3-sphere in $R^4$.  In
 addition,
the center of mass of the string may be rotating
 along another circle of $S^5$,
leading to  a particular string   solution
 with
all the three  $S^5$ charges  being non-zero $(J_1=J,\  J_2= J_3
 =J')$. The point-like  string case of \bmn\ corresponds to the
 special case  of $J'=0,\ J\not=0$  when the energy
of unexcited string is $E=J$.
In another special  case of $J=0,\ J'\not=0$ when the string has
 maximal
size  (so that $2J' \geq \sql$) the energy  turns out to depend
 on
$J'$ in a remarkably simple way: $E= \sqrt{(2J')^2 + \l} $.
While this solution with $J=0$ appears to be unstable,
there is always a non-trivial region of stability  when
$J\not=0$. As we shall show below, the solution is stable
 in the case  which is the  most interesting from
 the
 point of view of the AdS/CFT  comparison -- when  both $J$ and
 $J'$
are  large compared to $\sql$.

Let us start with a brief review of the basic features of this
 classical solution \ft.
Written in terms of the $AdS_5$ time coordinate   $t$ and
the angles of $S^5$ metric
 \eqn\Sd{
(ds^2)_{S^5}
= d\g^2 + \cos^2\g\ d\vp_1^2 +\sin^2\g\ (d\psi^2 +
\cos^2\psi\ d\vp_2^2+ \sin^2\psi\ d\vp_3^2)\ , }
the solution is
\eqn\ann{  t= \k \ta \ , \  \ \ \ \g=\g_0 \ , \ \
\ \ \vp_1= \nu \tau \ , \ \ \
 \ \ \vp_2 =  \vp_3 =\ww \tau \ ,  \ \ \ \ \  \psi= \kk \s \ , }
 where $  \k,\g_0,  \nu,\ \ww$ are constants,  $\kk$ is an
 integer\foot{
The standard range  of the angles $\psi,\vp_2,\vp_3$ to cover
$S^3$ only once is  $ 0 <\psi \leq {\pi\ov 2}$, $0  < \vp_2
\leq 2 \pi$,
$0  < \vp_3 \leq 2 \pi$:  then the
$S^3$\  ($\g = {\pi\ov 2})$  embedding coordinates
$X= \cos \psi\ e^{i\vp_2} , \  Y= \sin \psi\ e^{i\vp_3} $
have
positive ``radial'' factors (their sign change can be
compensated  by $\vp_{2,3} \to \vp_{2,3} + \pi$).
However, in the present case it is useful to choose
a different range: $ 0<  \psi \leq {2\pi}$, $0  < \vp_2
\leq  \pi$, $0  < \vp_3 \leq  \pi$.
Then the constant $\vp_2,\vp_3$ section  of $S^3$ will be the
 full
circle (instead of  its  $(0,{\pi\ov 2})$ quarter),
and thus we will have a consistent map of the closed string
into $S^3$ for {\it each} given moment of time
 $\tau$, as required by the  closed string interpretation.}
and $\tau$ and $\sigma \in (0,2\pi)$ are the
world-volume  (2-cylinder)   coordinates.
The equations of motion  and the
conformal gauge  constraints
imply
\eqn\relS{
\ww^2 = \nu^2 +\kk^2\ ,\ \ \ \ \ \ \ \ \
\k^2 = \nu^2 + 2\kk^2 \sin^2\g_0\ , }
so that  there are only  two continuous independent
 parameters -- $\nu$ (or $\k$)  and $\g_0$,
and one discrete one -- $\kk$.

The case of $\kk=0$ corresponds to the point-like solution
 considered in
\bmn\ and interpreted in the context of semiclassical
 approximation
in \rf{\gkp,\fts}. Then $\vp_1=\vp_2=\vp_3= \nu \tau$, and there
 is an
$O(6)$ rotation that transforms this solution into a null
 geodesic
along a canonical large circle of $S^5$.

In what follows we shall be mostly interested in
the minimal-energy  sector with  $\kk=1$
but will keep the $\kk$ dependence in some expressions for
 generality.

The non-zero  $SO(6)$ angular momentum components   $J_{MN}$
then are $J_1=J_{12}$, $J_2=J_{34}=J'$, $J_3=J_{56}=J'$, where
\foot{Here  it is assumed that $\kk \not=0$;
otherwise $J'$ is to be multiplied by 2.}
\eqn\sert{
J =\sql \nu \  \cos^2\g_0\ ,\ \ \ \ \ \ \
J'=  \ha \sql \ \sqrt{\nu^2 + \kk^2}\ \sin^2\g_0  \ .  }
$\sql={R^2\ov \a'}$ is the effective dimensionless
string tension related to
`t Hooft coupling.
The classical
 energy  $E=\sql \k= \sql \E_0(\nu,\g_0) $  is then
 $E=E(J,J',\lambda)$. It  can be expressed in terms of
the R-charge $ J'$ and the auxiliary ``charge''
 ${\II} = \sqrt{\l}\ \nu= {1 \ov \cos \g_0} \ J$:
\eqn\gtE{
E= \II \sqrt{  1 + {2\l \kk^2 \ov \II^2}
(1-{ J\ov \II })}  \ , }
where $\II=\II(J,J',\l)$ is a solution of the  (quartic) equation
\eqn\nuJ{
\II = J +   {2J' \ov  \sqrt{  1+{\l \kk^2 \ov \II^2} } } \   .
 }
Note that at the classical level
the dependence on $\kk$ can be absorbed into the
string tension $\sql$.

As we shall see below, at large $\nu$,
this solution  is stable under small perturbations provided
the spins  are subject to a certain condition; for example, 
for $\kk=1$, one needs to require 
\eqn\stbb{
J' \leq {3 \ov 2 } J \ , \ \ \ \ \ {\rm i.e.} \ \ \ \ \ \
{J' \ov   J + 2 J'} \leq { 3 \ov 8}   \ .   }
In the limit we will be interested in here
 when $\nu \gg \kk$  (similar to the  limit  considered  in
\bmn), i.e. when
${\sql  \kk\ov \II} \ll 1 $,  the
expression for the energy may be formally written as an
  expansion in positive powers of $\l $. More  precisely,
in the semiclassical approximation
one has,   of course,  $\l \gg 1$, but the  expansion is  in
${\l \kk^2  \ov \II^2 }\ll 1 $.

 Indeed, for large $J + 2 J'$  we can find
 $\II$  from \nuJ\
 as a series
in $\l \kk^2\ov ({J+2J'})^2$,  i.e.
\eqn\vev{ \II = J + 2J' - {\l \kk^2 J'\ov  (J+ 2J')^2} +
 ...\
  , }
so that
\eqn\rty{
E = J + 2J' + {\l\, \kk^2 J'\ov (J+ 2J')^2} + ...\ , }
where  the only requirement on   $J$ and $J'$
is  that $J + 2J'\gg\sql \ \kk$.
In addition to $\nu$ or $J + 2 J'$, the classical energy and
 quantum corrections
to it depend also on another parameter -- $\g_0$, which in the
$\nu \gg \kk$ limit is given  by (cf. \sert,\vev)
\eqn\rmm{
\sin^2 \g_0 =  { 2 J' \ov J + 2 J'} + ...   \ .  }
Note also that for   $J\gg J'$ the energy \rty\  becomes
\eqn\taal{
E = J + 2J'(1 + {\l\, \kk^2 \ov 2 J^2}  + ...)  \ , }
 which
 is consistent with  the  string oscillation
spectrum  in the sector with large
$J\gg\sql \kk  $ \rf{\bmn,\mets},
i.e. with the ``plane-wave'' \papa\ spectrum, where    $J'$
represents  the angular  momentum carried by string
 oscillations
(similar expression is found if $J'$ is replaced by spin in
$AdS_5$ directions).

In \ft\  it was  suggested that the  corresponding dual
${\cal N}=4$ SYM operators
should be of the form
$\tr [ (\Phi_1 + i \Phi_2)^{J} (\Phi_3 + i
\Phi_4)^{J'} (\Phi_5 + i\Phi_6)^{J'}]+ ...,  $ where dots stand
 for appropriate permutations of all $J = 2 J'$ scalar
 factors. These  operators belong to the irreducible
representation of $SU(4)$ with Young tableau  labels
$(J,J',J')$ or with Dynkin labels  $[0,J-J',2J']$ if $J\geq J'$,
and to the representation $(J',J',J)$ or $[J'-J,0,J'+J]$
if $J'\geq J$. 
We do not know if the solution with $\kk=1$ should be dual to an
operator having minimal canonical dimension for given values of
R-charges $J$ and $J'$.\foot{ The same formula \rty\ should be
giving  the conformal dimensions of the operators from the
two different ($[0,J-J',2J']$ or $[J'-J,0,J'+J]$)
representations. This  should be true  not only in
the large $\l$ limit but also  in
the weak-coupling  perturbation theory.}
There might exist a more complicated solution with less energy
describing, for example, a rotating folded string lying entirely
in $S^5$.

In the large $N$ SYM perturbation theory ($\l \ll 1$)
one expects to find  corrections to the canonical dimension
of these operators behaving  as
\eqn\wea{
\D(J,J',\l)_{\l \ll 1 }
  = J +   2 J' +  \l  F_1( J, J') +  O(\l^2)  \  .  }
The semiclassical result \rty\ is  a
prediction for the  anomalous dimensions
 in the opposite $\l \gg 1$
limit   when $J + 2 J' \gg 1$.
The dependence of the energy \rty\
on $\kk$ may be  reflecting  a  band structure of anomalous
 dimensions
of the SYM side.
\foot{These string solutions suggest that operators in  these
representations may be divided into different sectors
parametrized by the integer $\kk$. In each sector there is
an operator with the lowest conformal dimension that should be
dual to the string solution,  with  other operators in the
sector dual to excitations near that classical string
solution.}

Given a  simple dependence of the energy $E$ in  \rty\ on $\l$
in the limit $J + 2 J' \gg \sql  \kk$,
 it was conjectured in \ft\
that the expression   \rty\
may be  valid also at  {\it small } values
of $\l$ if $J + 2J'$ is very large.
More explicitly, one may expect  that the general  expression for
the  anomalous dimension  valid for any $\l$  but
with $J + 2 J' \gg \sql \kk$ (i.e. ${ \l\, \kk^2 \ov (J + 2 J')^2
 }
\ll 1$)
is
\eqn\clas{
\D(J,J',\l)_{_{J + 2J'\gg \sql \kk}}
  = J+  2 J' + \   f_1( J')    { \l \kk^2  \ov (J + 2  J')^2 } +
O( {\l^2 \kk^4 \ov  (J + 2J')^4 }) \    ,  }
where  in the string perturbation theory limit ($J' \sim \sql
 \gg 1$)
 \eqn\inte{
 f_1 (J')
 = J'   + {b_1} +  O( { 1 \ov J'})  \ . }
Similar expression is then expected for $\l \ll 1 $, i.e. in  the
 SYM
 perturbation theory.

In addition to the limit $J + 2 J' \gg \sql\gg 1 $,
another special case is   $J\gg J'$,   where one may  relate the
 resulting
 expression
for the energy  to the (non-perturbative)
corrections to the
dimensions of particular  operators in the  sector studied in
 \bmn.

On  general grounds, the   string sigma model corrections
to the  classical energy will have the following structure
\eqn\stru{
E=\sum_{l=0}^\infty E_l=  \sql \E_0 (\nu,\g_0) +  \E_1( \nu,\g_0)
 + { 1 \ov \sql } \E_2
 ( \nu, \g_0)   + ... \ , }
where $\E_1,\E_2,\E_3,...$  depend  only on the  parameters
$\nu,\g_0$ (and $\kk$) of the classical solution and should have,
for $\nu \gg \kk$,  an expansion  in { inverse}
 powers of $\n$, i.e.
in powers of $ \sql \kk \ov J + 2 J'$. For example,
\eqn\asd{
\E_0 (\nu,\g_0) = \k = \nu  + {1 \ov \nu } { \sin^2 \g_0 }
 + O( { 1 \ov \nu^2})  \ ,   }
\eqn\sedd{
\E_1 = \sum_{m=-1}^\infty {1\ov \nu^m} e_{m} (\g_0)\ ,  \ \ \ \ \
\ \ \ \ \  e_m = c_m + d_m \sin^2 \g_0  + ... \ .  }
Expressing this in terms of $J+2 J'\gg 1 $ and
${J'\ov J+2 J'}$ using \sert,\rmm\ we find that
the correction would be of the form \clas, \inte\ if:
  (i)
 the
functions $\E_1,\E_2,\E_3,...$ vanish in the limit
$\nu\to \infty$;
 (ii)  the expansion of $\E_1$  goes over only
even powers of $1/\nu$  and starts with $1/\nu^2$;
the expansion of $\E_2$  goes over only
odd powers of $1/\nu$ and starts with $1/\nu^3$, etc.
In particular, the  nonrenormalization of the leading $J+2J'$
 term
in $E$ requires the  vanishing the first three terms ($m=-1,0,1$)
in \sedd.\foot{Given that  for $\nu \gg \kk$  one has
${\sql \kk \ov J + 2 J'} \sim {\kk \ov \nu}  \ll 1 $  as another
small parameter of semiclassical expansion
(in addition to  ${ 1 \ov \sql}$),
 one may  expect, by analogy with
 the reasoning in \rf{\ft,\tse},    that  the leading $J + 2 J'$
 term in $E$
 will not  be renormalized to all orders in $1 \ov
 \sql$. Corrections
may  be suppressed in the large 2-d mass (large
 $\nu$) limit on a
 2-d cylinder. Note that all parameters in
 \ann,\relS\ scale as $\nu$ at large $\nu$ and there should be no
parameter-independent constant terms  in $\E_i$  because of
 supersymmetry.}
In that case all string corrections  can be written as functions
 of $J,J'$ and $\l$ with
$\l$ entering only in {\it positive} powers.
This is very similar to what was found  in the case of
 $\nu\not=0,  \
\kk=0$  \ft, i.e.  in the BMN case.
Assuming
that  the expansion of $\E_1$ starts with the $1/\nu^2$ term
(as we shall indeed confirm  below), one concludes that
the coefficient $b_1$ in \inte\ is subleading at large $J'$, and
is  equal
to $e_2(\g_0)$ in \sedd.
Indeed, we then  have
\eqn\ssss{
\E_1 = {1  \ov \nu^2}e_2(\g_0) + { 1  \ov \nu^4}e_4(\g_0) + ...
=  \bar e_2( {J'\ov J})  { \l \kk^2  \ov (J + 2  J')^2 }
+  \bar e_4( {J'\ov J})  { \l^2  \kk^4  \ov (J + 2  J')^4 } +
.. \ . }
Similarly, we expect that higher-order terms in \stru\ will have
 the structure
$$
E_l = {1 \ov (\sql)^{l-1}} \E_l ( \nu, \g_0)
= {1 \ov (\sql \nu )^{l-1}} [ {1  \ov \nu^2}e_{2l}(\g_0) +
 { 1  \ov \nu^4}e_{4l} (\g_0) + ... ] $$
\eqn\exx{
= { 1 \ov (J+ 2 J')^{l-1} } \bigg[ \bar e_{2l}
( {J'\ov J})  { \l \kk^2  \ov (J + 2  J')^2 }
+  \bar e_{4l} ( {J'\ov J})  { \l^2  \kk^4  \ov (J + 2  J')^4 }
 +
.. \bigg]\ . }
Thus, we expect that for large $J$
and $J'$ the classical expression for the energy \gtE\
will be giving the leading contribution
at any order in $\l$, and thus should be representing  also the
expression of the conformal dimension of the dual SYM operator
computed in a  weak-coupling expansion.

That  would  mean, in particular, that
the leading
one-loop perturbative correction to the dimension of the
corresponding CFT
operator dual to the string solution should indeed be of  order
$J'\ov {(J+2J')^2}$ and not  of  order $J+2J'$.
The non-renormalization of the leading $J$-term  in $E=\D$
does take place for the ground state  in the BMN  ($J'=0$) case,
where one expands near a point-like
 BPS state. In the present case of extended rotating string
 solution
 the space-time supersymmetry is broken,
but one may expect that it is in some sense
 ``restored'' in the limit $J + 2 J' \gg
\sql \kk $ (when a closed rotating string
is moving fast along a circle in $S^5$);
 that should then be   an explanation  for  the
 non-renormalization
 of the leading $J + 2 J'$ term in the energy.

\bigskip

After a review of classical solution in terms of the embedding
 coordinates
in Section 2,
 we shall derive    the
general expression for the 1-loop corrections to  the energy
 \rty\
coming both from the bosons and the Green-Schwarz fermions
 (Section 3).
We shall find the quartic equations for the (squares of)
characteristic  bosonic and fermionic frequencies.
In Section 4 we shall study the large $\nu$  (or, equivalently,
 large $\k$)  limit of the 1-loop correction
and confirm the structure of the 1-loop correction \ssss,
computing the value of $e_2(\g_0)$.
Some technical details will be   given in Appendices A,B and C.

The relevant bosonic and fermionic
 quadratic fluctuation parts of the Green-Schwarz
 \adss action were already presented in \ft\  and will
 be reviewed and further simplified  below.
In spite of the classical solution being dependent on $\tau$
and $\s$,
the quadratic fluctuation action can be
put (after natural local ``rotations'' of fluctuation fields)
in the form  where it describes a collection of
bosons  and
fermions  in flat 2 dimensional space  all having
constant  masses  and coupled to  constant (non)abelian 2-d
gauge terms.\foot{The constant connection terms
can    be eliminated at the expense of making the mass terms
non-diagonal and $\tau$  and $\s$-dependent.}
Remarkably, the form of this
  action is essentially the same  as of the
  light-cone gauge
superstring action in a   particular  plane-wave background
with an
antisymmetric 2-form field.
The problem of finding  the
leading correction to the ground state energy and  also of
the spectrum of string excitations near the
three-spin solution is thus closely related to the
corresponding problem in the case of the ``homogeneous''
 plane-wave backgrounds (cf. \rf{\plan,\papd}).

\newsec{Classical solution and bosonic part of
quadratic  fluctuation action}

Written in terms of the 6  embedding coordinates
of $S^5$ into $R^6$ (here we rename the coordinates $5,6 \to
1,2$)  \
($X_1^2 + ... + X^2_6=1 $) \ft\
\eqn\tikk{
 X=X_1 + i X_2 =\sin   \g \ \cos \psi \
e^{ i \vp_2} \ , \ \ \ \ \ \ \
Y= X_3 + i X_4=      \sin   \g \ \sin \psi \
e^{ i \vp_3} \  ,  }
$$ Z= X_5 + i X_6=\cos  \g \ e^{ i \vp_1} \ , $$
 the solution \ann\ is
\eqn\tokk{ \
   X= \sin \g_0 \  \cos \kk\s \ e^{ i \ww \ta} \ , \ \ \ \
Y=   \sin \g_0 \   \sin \kk\s  \ e^{ i \ww \ta } \  , \ \ \ \ \
\ Z= \cos \g_0 \ e^{i \nu \tau}  \ .    }
The conformal gauge constraints are then satisfied  provided
the coefficient $\k$ of the $AdS_5$ coordinate $t$ is related to
the parameters as in \relS.

This solution  can be found directly by starting
with the $S^5$ (or $O(6)$) sigma model
 action in conformal gauge
  ($\L$ is a Lagrange
multiplier  field, and $\eta_{ab}=(-,+)$)
$$ I_{S^5} = { \sql \ov 4 \pi } \int  d \tau \int^{2 \pi}_0
d \s \ L \ , \ \ \ \ \ \ \ \ \sql = {R^2 \ov \a'} \ ,  $$
\eqn\laga{  \ \ \ \
L = -\del_a X_m \del^a X_m    - \L ( X_m X_m - 1 ) \ , \ \ \ \ \
 \ \  \ \ \ \ \
\   m=1,...,6
\ . }
The classical equations of motion then are
\eqn\qqq{
 ( -\del^2   + \L ) X_m =0 \ , \ \ \ \
  X_m  X_m =1 \ , \ \ \   \ \ \
    \L = - \del_a X_m \del^a X_m  \ . \ \ }
They are satisfied  by \tokk\  with
\eqn\prov{ \ww^2 = \n^2 + \kk^2  \ ,\ \ \ \ \ \ \     \ \   \ \
\ \ \ \L= \nu^2  \ . }
This is an example of a  special class of simple
solutions of the non-linear equations \qqq\ for which
$\L=\const$ and  $X_m(\tau,\s) $ can be represented as a product
of commuting  $O(6)$ transformations depending on $\tau$ or $\s$
and applied to a  constant unit 6-vector.
Indeed,
we can write \tokk\ as ($\X \equiv  (X_m)$)
\eqn\wri{
\X (\tau,\s) =  O_{12+34}(\ww \tau)\ O_{13+24}(\kk\s)\
O_{56}(\nu\tau)\ O_{15}({\textstyle{\pi \ov 2}}-\g_0 )\  \X_0
\ ,
 \ \ \ \ \X_0 = (1,0,0,0,0,0) \ , }
where
\eqn\rott{
O_{pq}(\a) = e^{- \a  \JJ_{pq}} =
I+ \JJ_{pq}^2(1- \cos \a )   - \JJ_{pq} \sin \a \ , \ \ \ }
$$
O_{pq+ kl}(\a) = e^{- \a (\JJ_{pq} + \JJ_{kl})}  \  ,
\ \ \ \ \ \    (\JJ_{pq})_{mn}= \d_{pm} \d_{qn} -  \d_{pn}
\d_{qm}  \ ,    $$ and  $\JJ_{pq}$ is a generator of rotation in
the $(pq)$ plane in the fundamental representation of $O(6)$.
Note that $[\JJ_{12}+\JJ_{34},\JJ_{13}+\JJ_{24}]=0, \
[\JJ_{12}+\JJ_{34}, \JJ_{56}]=0, \     [\JJ_{13}+\JJ_{24},
\JJ_{56}]=0,$ i.e. the non-constant rotations commute. This
simplifies dramatically the form of the small-fluctuation action.

In general, the quadratic fluctuations near a solution of \qqq\
 are described by ($X_m \to X_m + \td X_m, \ \L \to \L + \td \L$)
  \eqn\fluc{
  L_2 =  -\del_a \td X_m \del^a \td X_m    -
   \L \td X_m\td  X_m   - 2 \td \L   X_m\td X_m \ ,  }
   i.e. they satisfy
    \eqn\takl{
    [\delta_{mn}   -   X_m X_n ]
   (- \del^2\td  X_n   + \L  \td X_n ) =0  \ , }
\eqn\conn{
\ \ \ \ X_m \td X_m=0 \ .    }
To find the action for the independent 5
 fluctuations we are thus to solve  the constraint \conn\
and substitute the result into the ``unconstrained''  action
\eqn\free{
 \td L_2 =  -\del_a \td X_m \del^a \td X_m    -  \L \td X_m\td
X_m   \  . } Finally, one  may
solve the (relevant linear part of) the conformal gauge
constraints, \eqn\conf{
- \k \del_\tau \td t + \del_\tau  X_m \del_\tau \td  X_m
+  \del_\s  X_m \del_\s \td  X_m = 0 \ ,\  \ \ \ \
- \k \del_\s \td t + \del_\tau  X_m \del_\s \td  X_m = 0 \ ,}
but this is not necessary in order to determine  the non-trivial
part of the spectrum. In addition,  one needs to include the
contribution of 4  massive bosonic fluctuations in the $AdS_5$
directions  \ft\ \eqn\sqw{
L_{AdS_5} = - \del_a y_l \del^a y_l - \k^2  y_l y_l \ , \ \ \ \ \ \ \
\ \ l=1,2,3,4 \ . }
In the present case of the solution \wri,\prov\  it is easy to
solve \conn\ (i.e. $ \X^T \td \X=0$)
by a field redefinition on $\td X_m$ that ``undoes'' the
rotation in \wri, i.e.
\eqn\wrw{
\td \X (\tau,\s) =  O_{12+34}(\ww \tau)\ O_{13+24}(\kk\s)\
O_{56}(\nu\tau)\ O_{15}({\textstyle{\pi \ov 2}}-\g_0 )
\  \bar \X  (\tau,\s)  \ , \   }
where  $\bar X_m $ are the   new (``tangent-space'')
fluctuations now subject to the simple $(\tau,\s)$-independent
condition $
 \bar X_m (X_m)_0 =0 $, which is solved  by setting
\eqn\zaz{    \bar X_1=0   \ .}
\subsec{Fluctuation Lagrangian}
Equivalently,
that means, combining \wri,\wrw,\zaz,
 that we parametrize
the classical+fluctuation field   as
$$\X (\tau,\s) =  O_{12+34}(\ww \tau)\ O_{13+24}(\kk\s)\ O_{56}(\nu\tau)\
O_{15}({\textstyle{\pi \ov 2}}-\g_0 ) \  \hat \X(\tau,\s)
\ ,$$
\eqn\clasa{
 \ \ \ \ \hat \X = (1,\bar X_2,  \bar X_3,\bar X_4,\bar X_5,\bar X_6) \ . }
Doing the transformation \wrw\ in   \free\
one ends
up with the following simple fluctuation Lagrangian with {\it
constant} coefficients (determined essentially by the generators
of $O(6)$ rotations in the classical solution):
$$ \td  L_2 =  (\del_\tau \bx_m )^2    - (\del_\s \bx_m)^2 $$ $$
 +  \ 4 \nu ( \cos \g_0 \bx_1 +  \sin \g_0\  \bx_5  ) \del_\tau
\bx_6 - 4 \ww [ ( \cos \g_0 \  \bx_5 - \sin \g_0 \bx_1)
\del_\tau \bx_2  -  \bx_3 \del_\tau \bx_4 ]
$$
 \eqn\flc{
+ \ 4 \kk[ ( \cos \g_0 \  \bx_5    -   \sin \g_0 \  \bx_1)
\del_\s \bx_3       -  \bx_2 \del_\s \bx_4]  \ , }
where  we have used integration  by parts.
Here the dependence on $\g_0$ could  be rotated away
if not for the constraint \zaz\  we still need to
impose. As a result, the fluctuation Lagrangian for the 5
independent fluctuation fields becomes ($s=2,3,4,5,6$)
$$  L_2 =  (\del_\tau \bx_s )^2    - (\del_\s \bx_s)^2
 + 4 \nu  \sin \g_0\  \bx_5   \del_\tau \bx_6
- 4 \ww (  \cos \g_0 \  \bx_5  \del_\tau \bx_2  -  \bx_3\del_\tau \bx_4) $$
 \eqn\fluca{
+ \  4 \kk( \cos \g_0 \  \bx_5     \del_\s \bx_3   -  \bx_2
\del_\s \bx_4)  \ .  }
The point-like (BMN) limit corresponds to the case of $\kk=0$
(then $\ww=\nu$). The resulting Lagrangian
can be shown to be equivalent to the one found
by expanding near a ``canonical'' BMN solution
$ L_2 = - (\del_a \bar X_5 )^2     - (\del_a \td X_i )^2 -
\n^2  \td X_i^2 $, where $i=1,2,3,4$ (with the  constraint \zaz \
now being $\bar X_6=0$).

Eq. \fluca\  is a special case of the following 2-d Lagrangian
\eqn\spacd{
  L =  (\del_\tau x_p )^2    - (\del_\s x_p)^2
 + 2 f_{pq}   x_p   \del_\tau x_q
 -  2 h_{pq}   x_p   \del_\s   x_q \ , }
where $f_{pq}$ and $h_{pq}$ are constant antisymmetric
coefficient matrices. The latter can  be written also as
(ignoring total derivative)
\eqn\spac{
 L =  (\del_\tau  x_p  + f_{qp} x_q )^2    - (\del_\s x_p   +
h_{qp} x_q)^2  - (f_{pq} f_{kq}  -  h_{pq} h_{kq}) x_p x_k \ , }
i.e. it represents a   massive scalar 2-d theory coupled to a
constant 2-d  gauge field
(which can be ``rotated away''  at the expense of
 making  the mass term $\tau$ and $\s$
dependent).
The corresponding Hamiltonian  is
\eqn\spac{
 H =  (\del_\tau  x_p)^2    + (\del_\s x_p   +
h_{qp} x_q)^2  
 - h_{pq} h_{kq} x_p x_k \ . }
 In the case of \fluca\ we find
$$  H_2 =  (\del_\tau \bx_2 )^2  +  (\del_\tau \bx_3 )^2  +
(\del_\tau \bx_4 )^2
+ (\del_\tau \bx_5 )^2 +   (\del_\tau \bx_6 )^2
  + (\del_\s \bx_2  - \kk \bx_4)^2
  + (\del_\s \bx_4  + \kk \bx_2)^2  $$ $$
  +\  (\del_\s \bx_3  - \kk \cos \g_0 \bx_5)^2
  + (\del_\s \bx_5  +  \kk\cos \g_0 \  \bx_3)^2
  $$
\eqn\zagh{  
- \ \kk^2 \cos^2 \g_0 \  (  \bx_3^2  +  \bx_5^2)
-  \kk^2 (\bx_2^2  +  \bx_4^2)
  \ . }
While the Hamiltonian is always positive in the point-like case
$\kk=0$, it is not manifestly so
 for $\kk > 0$, i.e. there  is a potential  for an instability.
In general, the conclusion about  non-positivity and instability is not directly obvious
on a cylinder; for example,  we cannot set
$\del_\s \bx_4  + \kk \bx_2 =0$ for constant $\bx_2$
since then $\bx_4$ will not be periodic in $\s$.
The instability  is always present when $ \g_0={\pi \ov 2}$,
i.e. when $\bx_6$ and $\bx_2$ are decoupled.
As we shall argue below, there is a range  of parameters
(large $\nu$ and sufficiently small $\g_0$) for which the  solution  is  stable
under small perturbations.

Let us note that the  Lagrangian \spacd\     can  be also interpreted
as a light-cone gauge  ($u=\tau$)  Lagrangian  for the
bosonic string sigma model
$L= - ( \eta^{ab} g_{mn} + \epsilon^{ab} B_{mn} ) \del_a x^m
\del_b x^n $  in  (in general, non-conformal)   plane-wave
background  with the following metric
and the antisymmetric 2-form field (cf. \papd)
\eqn\ppw{
ds^2 = 2 dudv   + 2 f_{pq} x_p dx_q du  +  dx_p dx_p  \ ,
 \ \  \ \  \ \ \ \
B_2  =  2 h_{pq} x_p  d x_q \wedge  d u  \  . }
The general form of (the linear part of) the
conformal gauge constraints is
$$
\k \del_\tau \td t =
\k^2 \bx_1  - \kk^2 \sin 2 \g_0 \  \bx_5 + \kk\sin \g_0 \
\del_\s \bx_3   + \ww \sin \g_0 \ \del_\tau y_1  + \nu \cos \g_0\
\del_\tau \bx_6
$$
\eqn\confi{
\k \del_s \td t=
2 \kk\ww \sin \g_0\  \bx_4   + \kk\sin \g_0 \ \del_\tau \bx_3
 + \ww \sin \g_0 \ \del_\s y_2  + \nu \cos \g_0\  \del_\s \bx_6
\ . }
Adding $- \del_a \td t \del^a \td t$ term and eliminating $t$
from the resulting action  implies cancellation also of one
(``massless'') combination of $\bx_s$ coordinates; after a field
redefinition one ends up with the following  ``reduced''
Lagrangian  for the remaining 4 non-trivial fluctuations:
\eqn\redd{  L'_2 =  (\del_\tau \bx_s )^2    - (\del_\s \bx_s)^2
- 4 \nu C_1  \bx_5   \del_\tau \bx_6 - 4 \k  C_2 \bx_5  \del_\tau
\bx_2  +   4 \k  C_1   \bx_4 \del_\s \bx_2 +   4 \nu  C_2
\bx_4     \del_\s \bx_6    \ ,  } $$ C_1  = [{  \kk^2 (1 + \sin^2
\g_0)   \ov \n^2  + \kk^2 \sin^2 \g_0 }]^{1/2}  \ , \ \ \ \ \
C_2  =  [{ \n^2 +  \kk^2    \ov  \n^2  + \kk^2 \sin^2 \g_0 }]^{1/2}
 \ . $$
In the special case of $\nu=0$ this is equivalent to the
fluctuation Lagrangian obtained in \ft\ in the static gauge.

\subsec{Characteristic frequences}

To find the spectrum  of characteristic frequencies
corresponding to the action  \fluca\ (for a general analysis of
the theory \spacd\ see \papd) we  note that $\bx_s$ fields  must
be periodic in $\s$ (the rotations \wrw\  we made preserve the
periodicity) so that  one can expand the solution of the
quadratic fluctuation equations  in modes
\eqn\zas{
\bx_s = \sum_{n=-\infty}^{\infty} \sum_{i=1}^8  A^{(i)}_{sn}
e^{i( \om_{n,i} \tau  \ + \  n \s)}  \ ,  }
where $i$ labels different frequencies  for a given value of $n$
(we shall often  suppress the index $i$ below).
Plugging this into the classical equations that
follow from \fluca\    one finds  the following
result  for the determinant of the characteristic matrix
$$ \det M =
-(n^2- \Om) B_8(\Om)   \ ,\ \ \ \  \ \ \ \ \ \ \  \Om  \equiv
\om^2_n \ ,  $$ $$  B_8(\Om)  =  \ \Omega^4
-  ( 6 \kk^2  + 8 \n^2  + 4 n^2  + 2 \kk^2 \cos 2 \g_0 ) \
\Omega^3   $$ $$ + \   [ 8 \kk^4 + 6 \kk^2 n^2  + 6 n^4  + 24
\kk^2 \n^2  + 16 n^2 \nu^2 +16 \n^4 + ( 8 \kk^4 + 2 \kk^2 n^2  +
8 \kk^2 \n^2) \cos 2 \g_0 ] \ \Omega^2 $$ $$ +\ [ -16  {\kk^4}
{n^2}+6 {\kk^2}{n^4} -4 {n^6} -8 {\kk^2} {n^2} {{\nu}^2} -  8
{n^4} {{\nu}^2}  + ( -16 {\kk^4} {n^2} +  2 {\kk^2} {n^4} -24
{\kk^2} {n^2} {{\nu}^2} ) \cos 2 \g_0] \Om   $$
\eqn\frt{   + \
n^4 ( n^2 - 4 \kk^2) (n^2 - 2 \kk^2 - 2\kk^2  \cos 2 \g_0 ) \ .
}
Setting $\det M =0$, we  observe the existence of one decoupled
massless scalar field corresponding to the solution $\Om = n^2$
of \frt. The decoupled massless scalar is a reflection of the
conformal gauge choice we made.
We also find a nontrivial quartic equation  for the
remaining modes, giving 4 (in general, different) values for
$|\w_n|$, i.e. 8 characteristic frequencies $\w_{n,i}$.  Here
$\kk$ can be set to 1 (it can be restored by $n \to n /\kk , \
\nu \to \nu /\kk, \ \om_n \to \om_n /\kk $).

 The same   result for the characteristic determinant (but
without the zero-mode factor) is found by starting from the
``reduced'' action \redd.

The BMN limit corresponds to setting $\kk=0$ in the above formulae;
while the fluctuation Lagrangian \fluca\  seems to depend on $\g_0$,
the spectrum, as one might  expect,
 does not:  for $\kk=0$ the determinant
\frt\  becomes
\eqn\dert{
\kk=0 \ : \ \ \ \ \ \ \ \
B_8= (n^2 - 2 \nu \w_n - \w_n^2)^2 ( n^2 + 2 \nu \w_n -
\w_n^2)^2  \ , } and thus the roots are
\eqn\aqw{
\w_n = \pm \nu \pm   \sqrt{ n^2 + \nu^2}  \ .  }
Here the linear $\nu$ terms reflect the rotation of the
fluctuations made in \wrw\  while $ \pm  \sqrt{ n^2 + \nu^2}$ are
the standard ``plane-wave'' frequencies. Similar result is found
also in the  fermionic sector discussed in the next section.

Another special case is when $\g_0$ is approaching $\pi\ov 2$
\eqn\sweq{
\g_0= {\pi \ov 2}: \ \ \ \ \ \  J=0\ ,\ \ \ \ J'\not=0  \ , \ \
\  \ \ \ \ \ E= 2J' \sqrt{1 + {\l \kk^2 \ov (2J')^2}} \ . }
Setting $\g_0= {\pi \ov 2}$  in \frt\ and solving $B_8=0$ we
find the following 4+4 frequencies
\eqn\freqq{
\w_n^2 = n^2 + 2 ( \nu^2 + \kk^2)
\pm  2 \sqrt{ (\nu^2 + \kk^2 )^2 + n^2 ( \nu^2 + 2 \kk^2) }  \ , }
\eqn\zazz{
\w_n = \pm \nu \pm \sqrt {n^2 + \nu^2} \ . }
The condition of reality of $\w_n$ in  \freqq\
is $ n^2(n^2 - 4 \kk^2)  \geq 0$,  i.e. this solution  has
unstable modes with $n=\pm 1,\cdots ,\pm (2\kk -1)$, as was
already found in  \ft\ for $\kk = 1$.

$J=0$  is found also when $\nu=0$ (see \sert);
in this case
\eqn\taks{
\nu=0 \ : \ \ \ \J' =  \ha \sql \kk \ \sin^2 \g_0
\ ,  \ \ \ \ \ \ \    J=0 \ , \ \ \ \ \ \
E= \sqrt{ 4 \sql \ \kk \ J'} \ ,  }
i.e. here $J'$ bounded from above
for fixed integer $\kk$ ($\kk=1$ case was discussed in \ft).
Here the non-trivial characteristic frequences are
\eqn\rte{
\w_n^2 = n^2 + 2 \kk^2 ( 2 - \sin^2 \g_0)
\pm 2 \kk \sqrt{ 2 n^2 (  2 - \sin^2 \g_0)
 + \kk^2 \sin^2 \g_0 }   \ . }
The condition of reality of $\w_n$
here is
$( n^2 - 4 \kk^2 ) ( n^2 - 4 \kk^2 + 4 \kk^2 \sin^2 \g_0)  \geq 0$,
which is
 satisfied
for
$\sin^2 \g_0 \leq { 4\kk -1 \ov 4 \kk^2} $, i.e.
$J' \leq  \sql { 4 \kk -1  \ov 8 \kk} $.
The same stability condition was found in \ft\ in  the special case
of $\kk=1$.
 Thus for $\nu=0$  the combination
$ { \l \ov J'^2} \geq ({ 8 \kk \ov 4 \kk -1})^2  $
cannot be made  small for any $\kk$.

For generic values of the parameters
 the  expressions for the frequencies $\w_{n,i}$  cannot be
written in a useful form,  but it is straightforward to
determine their form in large $\nu$ (or, equivalently,  large
$\k$) expansion. The results are presented in Appendix A.

The quartic equation $B_8=0$ leads  to 4
solutions for $\om_n^2$. If all of them are non-negative, the
solution is stable. Let us analyse the stability condition in
the large $\k$ limit.
 In this limit there are two different
asymptotics of the different  frequencies:
\eqn\zdw{
(i) \ \ \ \om_n^2 \to {h_0 \ov 4 \k^2}  + {h_1 \ov \k^4} + ...
\ , \ \ \ \ \ \ (ii) \ \ \ \om_n^2 \to {4 \k^2}  + g_0 + {g_1
\ov \k^2} + ...   \ . }
One finds  (here  we set $\kk=1$ and  $n \geq 0$):
\eqn\sas{
h_0 = n^2 \bigg[ n^2 + 4 - 6 \sin^2 \g_0
\pm 2 \sqrt{ 4 n^2 \cos^2 \g_0  - 8 \sin^2 \g_0  + 9 \sin^4 \g_0
}\bigg]
\ . }
The condition of non-negativity of $h_0$ is obtained at $n=1$, and
is $\sin^2 \g_0 \leq { 3 \ov 4}$.\foot{Note that the limiting  case
$  \sin^2 \g_0 = { 3 \ov 4}$
corresponds to the value of the angle $\g_0 = {\pi  \ov 3} $ and
may have some  geometrical interpretation.}
 This implies the stability condition \stbb.

In general, for $\kk^2 > 1 $ 
(the analog of \sas\ in this case is found  by 
replacing $n \to {n\ov \kk}$ in the bracket 
in \sas\ and adding overall factor of $\kk$)  
the stability condition is obtained 
by requiring that 
$ (q^2-4) (q^2 - 4 \cos^2 \g_0) \geq 0 $ as 
well as 
$ (3 \cos^2 \g_0 -1)^2  + 4 \cos^2 \g_0\ ( q^2-1)\geq 0 $, where we set
$q= {n \ov \kk}$. 
It is straightforward to show that for each value of $q$
there is a range of values of $\cos \g_0$ where these conditions are
satisfied.


In the  case (ii) one finds
\eqn\dedp{
\om_n = 2\k  + { 1 \ov  2\k} [ n^2 + 1 - 5\ss \pm \sqrt{4 n^2
\cos^2\g_0 + \sin^4 \g_0 }] + ...  \ .
}
For comparison, a similar expansion of the frequencies
$\sqrt{\k^2 + n^2}$ in the $AdS_5$ directions in \sqw\ is
 $$ \om_n = \k  +  {n^2  \ov 2 \k}
-  {n^4\ov 8 \k^3 }   + ... \ .   $$

\newsec{Fermionic part of the quadratic fluctuation action}

Let us  first recall the basic  expressions
for the quadratic fermionic action (see Appendix B in \ft) and
then find the corresponding spectrum of fluctuations.
The quadratic part of the type IIB
\adss  Green-Schwarz superstring action
expanded near a particular bosonic string solution
(with flat induced metric)   is
\eqn\fer{
L_F=
i (\eta^{ab }\delta^{IJ} -
\ep^{ab } s^{IJ} ) \bar \vt^I \vr_a D_b \vt^J   \ ,\ \ \ \ \ \
\vr_a \equiv \G_{A}  e^A_a \ , \ \ \  e^A_a\equiv E^{A}_M(X)
\del_a X^M \ , }
where
 $I,J=1,2$,  $s^{IJ}=$diag(1,-1),
  $\vr_a$ are projections of the 10-d Dirac
matrices and $X^M$ are the  string coordinates
corresponding to the  $AdS_5$  ($M=0,1,2,3,4$) and $S^5$
($M=5,6,7,8,9$)
factors.
The covariant derivative $D_a$ is
\eqn\form{
D_a\t^I   = (\delta^{IJ} {\rm D}_a
- { i \ov 2 } \epsilon^{IJ}  \G_* \vr_a ) \vt^J\ ,
\ \ \ \ \ \ \ \ \  \G_* \equiv i \G_{01234} \ , \ \
\G_*^2 =1 \ ,
 }
 \eqn\topp{ {\rm D}_a = \del_a
+\fourth    \omega^{AB}_a\Gamma_{AB} \ ,
\ \ \ \ \ \ \ \omega^{AB}_a \equiv   \del_a  X^M \omega^{AB}_M
\ .}
Choosing the $\k$-symmetry gauge by
equating the two Majorana-Weyl 10-d
spinors,
\eqn\ghh{ \vt^1=\vt^2 \equiv \vt \ , }
  we get
\eqn\fert{
L_F=
2i \bar \vt  D_F  \vt  \ , \ \ \ \ \ \ \ \  D_F=  - \vr^a \rD_a
 - {i\ov 2} \epsilon^{ab} \vr_a  \G_* \vr_b  \ .  }
 In the case of the $S^5$ solution \ann\
 with $\kk=1$
 we shall label the tangent space coordinates
 by  $A=0,5,6,7,8,9$  corresponding to the
$t$ direction of $AdS_5$ and $\g,\vp_1,\psi,\vp_2,\vp_3$
directions of $S^5$.
 Then  \ft\
  \eqn\varr{
 \vr_0 = \k  \G_0 +  \n \cos \g_0 \ \G_6  + \ww \sin\g_0 \ \td
 \G_8\ , \ \ \
 \ \ \ \vr_1 =
 \sin \g_0 \ \G_7 \ ,  \ \ \ \ \ \
  \ \   \vr_{(a} \vr_{b)} = \sin^2 \g_0
\ \eta_{ab} \ ,  }
\eqn\vrua{ \td \G_8 \equiv  \cos \s \ \G_8  +  \sin \s \ \G_9 \ , \
\ \ \ \ \
 \td \G_9 \equiv  \cos\s\ \G_9   -    \sin\s \ \G_8 \  . }
The projected Lorentz connection
has the following non-zero components
$$\omega^{65}_0 = -\n \sin\g_0  \ ,  \ \ \ \   \ \ \
\omega^{85}_0 = \ww\cos\g_0\   \cos \s  \ , \ \ \ \ \ \
\omega^{95}_0 = \ww \cos \g_0\ \sin\s  \ , $$
  \eqn\conm{\ \ \
\omega^{87}_0 = -\ww \sin \s  \ , \ \ \ \ \ \
\omega^{97}_0 = \ww \cos\s  \ , \ \ \ \ \ \
\omega^{75}_1 = \cos \g_0 \ .  }
To eliminate the $\s$ dependence  we  shall
 first do a local  rotation in the 89-plane:
\eqn\rott{ \vt = S(\s) \td \vt, \ \  \ \ \ \
 S = e^{ - \ha \s \G_{89}} \ , \ \ \ \ \
  S^{-1}  \td \G_i S = \G_i  \ , \ \ \ \  i=8,9 \ . }
As a result, $\td \theta$ will be {\it anti}periodic in $\sigma$.

 Then we get for the fermionic   operator in \fert\
$$
D_F =  ( \k  \G_0 +  \n \cos \g_0 \ \G_6  + \ww \sin\g_0 \
\G_8) ( \del_\tau  - \ha \nu \sin \g_0\ \G_{65} + \ha   \ww
\cos \g_0\ \G_{85} + \ha \ww \G_{97} )   $$  \eqn\traa{
- \ \sin \g_0\  \G_7 ( \del_\sigma  + \ha \cos \g_0 \ \G_{75}  -
\ha \G_{89} ) +  \sin \g_0 \G_7 (  \n \cos \g_0 \ \G_6  + \ww
\sin\g_0 \ \G_8) \G_{01234} \ .  }
We can put \traa\ in the form
$$
D_F =  ( \k  \G_0 +  \n \cos \g_0 \ \G_6  + \ww \sin\g_0 \ \G_8)
\del_\tau - \ \sin \g_0 \G_7  \del_\sigma
$$
\eqn\fre{
+ \  \ha \G_5 ( - \nu \k \sin \g_0 \ \G_{06}  +
\ww \k \cos \g_0 \ \G_{08}  + \nu \ww \G_{68} )
}
$$
+\   \ha [ \k \ww \G_0 +  \n \ww \cos \g_0 \ \G_6  +( \ww^2 + 1)
\sin\g_0 \  \G_8] \G_{79}  $$ $$ -\    \sin \g_0  (  \n \cos \g_0
\ \G_6  + \ww \sin\g_0 \ \G_8) \G_{07} \G_{1234} \ .  $$
This can be simplified further  by making two constant rotations
in the $68$ and $06$ planes  to
transform $ \k  \G_0 +  \n \cos \g_0 \ \G_6  + \ww \sin\g_0 \
\G_8$ into $ \sin \g_0 \G_0$:
 \eqn\anot{
 \td \vt =  S_{68} S_{06}    \Psi\ , \ \ \ \ \ \ \
 S_{68}  = e^{ -\ha p \G_6\G_8} \ , \ \ \ \
 S_{06}  = e^{  - \ha q \G_0\G_8} \ , \ \ \ \  }
\eqn\serf{
 \cos p = {\n\ov a }  \cos \g_0\   , \ \ \
 \sin p = {\ww\ov a }  \sin \g_0 \  , \ \ \ \
a\equiv  \sqrt{ \n^2 + \sin^2 \g_0}  \ , }
$$
 \cosh q = {\k\ov \sin \g_0 } \  , \ \ \
\ \ \    \sinh q = {a \ov  \sin \g_0 } \  . \
$$
Then rescaling $\Psi$ by  $(\sin \g_0)^{1/2}$
 we  finish with the following fermionic Lagrangian
 $$
L_F= 2i \bar \Psi  \big[  - \G_0 ( \del_0  +
 { \k \cos \g_0 \ov 2a} \G_{65} +  { \ww \nu \ov 2a} \G_{85} )
$$\eqn\ferk{
+ \  \G_7 (\del_1    -    { \k \ww \ov 2a} \G_{69}  -
{ \nu \cos \g_0 \ov 2a}  \G_{89}) -  a  \G_{07} \G_6 \G_{1234}
\big] \Psi \ ,   } or
\eqn\frk{
L_F=2i \bar \Psi  \big[  \tau^0 (\del_0 + A_0)   +
\tau^1 (\del_1  + A_1 ) -  a  \tau_3  \G_6 \Pi  \big] \Psi  \ , }
where
\eqn\ghj{   \tau_a\equiv  (\G_0, \G_7) \ , \ \ \  \tau_3 =
\tau_0 \tau_1 \ , \ \ \  \  \Pi= \G_{1234} \ , \  \ \ \Pi^2 = I \
,  } \eqn\zdd{ A_0 = { 1 \ov 2 a } (  \k \cos \g_0 \G_6 + \ww \n
\G_8)\G_5    \ ,\ \ \ \ \ \ \ \ \ A_1  =- { 1 \ov 2 a } (\k \ww
\G_6 +   \nu \cos \g_0 \G_8 ) \G_9  \ . }
 This action may be interpreted
as describing
a collection of eight standard  2-d  massive  Majorana fermions
 on a {\it flat}  2-d background
coupled to a constant non-abelian 2-d  gauge field $A_0,A_1 $.
We may also  split  the  fermions into
the eigen-states of the projector
$P= \ha ( I + \G_{1234})$
(which commutes with the rest of the operator).
 If one  chooses a  representation for  $\G_A$ where
$\G_0$ and $\G_7$ are  2-d Dirac matrices  times
a unit $8\times 8$
matrix  one  gets
4+4 species of 2-d Majorana fermions  with masses  $\pm a=\pm
\sqrt{\nu^2 + \sin^2 \g_0}$.

Note that in the large $\nu$ limit $\k,\ww,a \to \nu$,  i.e.
\eqn\zde{ A_0 \to
 { 1 \ov 2  }  \n \G_8\G_5    \ ,\ \ \ \ \ \ \ \ \
A_1  \to - { 1 \ov 2  }\nu  \G_6  \G_9  \ ,
}
so the action \frk\   simplifies.

 While the presence of the  $\tau_3 $ mass term  in \ferk\
 has its origin in the coupling of the GS fermions
 to the  5-form background \MT,  the presence of the $A_a$
 connection term in \frk\  may be also  interpreted as been
  due to the coupling to an effective NS-NS
   background in \ppw.
 For example, the coupling of the GS fermions to $H_{MNK}$
 field strength is through the  term in the covariant derivative
 $ D_a \vt^{1,2}  =
  (\del_a \pm  { 1 \ov 8}  \del_a X^K  H_{KMN} \G^{MN} + ...)
  \vt^{1,2}  $.  In the gauge \ghh\ the $H_{KMN}$ term
  contributes through the $\epsilon^{ab}$ term in \fert,
  i.e. we get   $ \sim \bar \vt \epsilon^{ab}
    \vr_a \del_b X^K  H_{KMN} \G^{MN} \vt $.
 For  $X^K\to  u=\tau$   we get   extra  term
 $ \sim \bar \vt   \vr_1  H_{uMN} \G^{MN} \vt$ which for the
background in \ppw\ produces terms in $A_1$ in \frk.
Thus expanding near a  circular  classical string  solution
induces an effective    $H_{MNK}$ background in both the bosonic
and the fermionic fluctuation    sectors.

To solve the Dirac equation corresponding to \ferk\ one should
 recall that while  the original GS spinor variable $\vt$ in \fer\
 was periodic on the 2-d cylinder,  the rotated fermion
 $\td \vt$ in \rott\  and thus $\Psi$ in \anot\  is  then
 anti-periodic, i.e. $\Psi(\ta,\s + 2 \pi) = -\Psi(\ta,\s )$.
That means one should look for solutions in the form
 \eqn\solf{
 \Psi = \sum_{r\in Z+\ha} \sum_{i=1}^8 \psi^{(i)}_r \  e^{i (
\omega_{r,i}  \tau \ + \  r \s)  } \   . }
The frequences $\w_r$ can be found by solving the characteristic equation
$F_8(\w_r) =0$. The latter
can be obtained 
 by  multiplying the Dirac operator
in \frk\ by its  appropriate ``conjugations'' or by
using an explicit representation for  6 Dirac matrices
$\G_0,\G_5,...,\G_9$\foot{Since we only need their algebraic
relations these $\G_m$
 can be taken as  Dirac matrices in 6 dimensions.}
 (here $\Omega=\omega_r^2$, \  $\kk=1$, cf. \frt)
$$
F_8= \Omega^4
- (5 + 4\  r^2 + 6\ \n^2   - \cos 2 \g_0 )\  \Omega^3 $$ $$
+\  \big[6 + 6\  r^4 + 14\ \n^2 + 9\ \n^4 +
2\  r^2  (2 + 7\ \n^2) +2 (1 + 5\ r^2 +
2\ \n^2)\ \sin^2 \g_0 + {3\ov 2} \sin^4
\g_0\big]\Omega^2 $$ $$
 - 2\ \big[ 2 + 2\ r^6 + 5\ \n^2 + 5\ \n^4 + 2\ \n^6 +
 r^4\ (-2 + 5\ \n^2) +
  r^2(-2 - 2\ \n^2 + 5\ \n^4)  $$ $$ + \
 [-1 + 7\ r^4 +  \n^4  +
 2 r^2  ( 1 + 6\ \n^2)]\ \sin^2 \g_0  - \ha
 (1 - 7\ r^2 +\ha  \n^2)\ \sin^4 \g_0  + { 1 \ov 4}  \sin^6
\g_0\big] \Omega  $$
$$ +\
( r^2-1)^2\ [r^4 +
                2 r^2 ( \n^2-1)  + (\n^2+1)^2]
  +
  2(  r^2-1)[3\ r^4 +  2\ r^2  ( \n^2-2) +(1 + \n^2)^2]  \sin^2 \g_0
$$ \eqn\zdf{
+
\ha [3 + 19\ r^4 + 5\ \n^2 + 2\ \n^4 +
r^2\ (-14 + 5\ \n^2)]  \sin^4 \g_0 +
\ha (-1 + 3\ r^2 - \n^2)\ \sin^6 \g_0 + { 1 \ov 16}
\sin^8  \g_0  \ . }
As in the bosonic case, while one cannot write down  simple
expressions for the frequencies in general,
one may expand  $\w_r$ at large $\nu$.
The results are presented in Appendix B.

The expressions for the frequencies take  simple form when
$\nu=0$  (there is 4+4 degeneracy):
\eqn\royu{
\nu=0: \ \ \ \ \ \
\w^2_r    = r^2 + 1  + \ha \sin^2 \g_0  \pm \sqrt{ 2 [ (2-
\sin^2 \g_0) r^2 +  \sin^2 \g_0  ]}  \ . }
Another special case is the one of the unstable solution \sweq\ with
 $\g_0= {\pi \ov 2}$  where we find
\eqn\daaa{\g_0= {\pi \ov 2} \ : \ \ \ \ \
\w_r = \pm \ha \big[  \nu \pm
\sqrt{\nu^2 + 2 }
\pm 2\sqrt{  r^2 + \nu^2 + 1  } \ \big]
\ . }
Let us mention also that to establish the connection to the
point-like (BMN) case we need first to restore the dependence of
$F_8$ in \zdf\ on the discrete
parameter $\kk$  (which can be  done by the rescaling
$ r\to r/\kk, \ \nu \to \nu/\kk, \
\Omega \to  \Omega/\kk^2 $,\  $F_8 \to  \kk^{8} F_8$)  and then
send $\kk$. As a result, we find  that $F_8$ becomes  (cf. \dert)
\eqn\derf{
\kk=0 \ : \ \ \ \ \ \ \ \
F_8 = (r^2 + \nu^2 - \w_r^2)^2
(r^2 - 2 \nu \w_r - \w_r^2) ( r^2 + 2 \nu \w_r - \w_r^2)  \ .  }
Here one is also to replace $r$ by $n$ taking integer values,
since in the $\kk=0$ limit there is no local rotation \rott\ of
the fermions that changes their periodicity in $\s$.
Thus, as for  the bosonic fluctuations \aqw,
the roots of $F_8=0$ are
indeed the same (up to a $\tau$-dependent rotation contribution)
as the   ``plane-wave'' frequencies.

\newsec{One-loop string sigma model  correction to the energy}

In this section we use the bosonic and fermionic spectrum to
compute the one-loop sigma model correction to the energy of the
solution. As in the static gauge $t=\k \tau$  in \fts,
the space-time energy and the 2-d energy (sum of $\ha \w$ for all
oscillator  frequencies) are related by
\eqn\zcv{   E = { 1 \ov \k} {\rm E}_{2-d} \ .   }
Then   the 1-loop correction is given by the  standard sum
of the oscillator frequencies sums
\eqn\corre{
  E= E_0+ E_1 + ...   \ , \ \ \ \ \ \
E_1 = {1\ov 2\kappa} \big(\ \sum_{n\in Z} \w_n^B -
\sum_{r\in Z+\ha} \w_r^F\ \big)\ ,}
where $\w_n^B$ and $\w_r^F$ are  bosonic and fermionic
contributions, respectively:
\eqn\correBF{
\w_n^B = \sum_{i=1}^8 \w_{n,i}^B \ , \ \ \ \  \ \ \ \ \ \
\ \w_r^F =
\sum_{i=1}^8 \w_{r,i}^F \ ,}
 where the index $i$ labels the characteristic frequencies.
This expression is UV finite, as one  can show   directly from
the expression for the  total fluctuation Lagrangian
\sqw,\fluca,\frk, or from the large $n$ and large $r$ expansions
of the frequencies given in Appendices  A  and B  (see also \ft).

One can check that  the 1-loop correction vanishes  in the
``point-particle'' limit  $\kk=0$, in agreement with the
non-renormalization of the energy of this  BPS state
dual to a gauge theory  operator with protected conformal
dimension \bmn. In what follows we shall set $\kk=1$.

We would like  to compute \corre\ as an expansion in $1\ov \k$
in the large $\k$ limit. In the large $\k$
limit there will be  also exponentially small terms which we
shall disregard.
To estimate the value of the sums we shall approximate them by
integrals as explained in Appendix C.

As discussed  in Appendices A and B, the bosonic and fermionic
frequencies
$\w_{n,i}^B$ and
$\w_{r,i}^F$ admit the following large $\k$ expansion
\eqn\omb{
\w_{n,i}^B = \k\, \hw_{-1,i}^B({n\ov \k}) + \hw_{0,i}^B({n\ov
\k}) + {1\ov \k}\hw_{1,i}^B({n\ov \k}) + \cdots \ ,}
\eqn\omf{
\w_{r,i}^F = \k\, \hw_{-1,i}^F({r\ov \k}) +
\hw_{0,i}^F({r\ov \k}) + {1\ov \k}\hw_{1,i}^F({r\ov \k}) +
\cdots \ ,} where we keep ${n\ov \k}$ and ${r\ov \k}$ fixed in
the expansion. The values $\hw_{a,i}({m\ov \k})$ can be
considered as values of the functions $\hw_{a,i}(x)$ at points
$x_m={m\ov \k}$, where $m\in Z$ for bosons, and $m\in Z +\ha$ for
fermions, and $\D \equiv x_{m+1}-x_m={ 1\ov \k}$.

We also need to regularize the bosonic and fermionic sums.
 This  can be done by
multiplying each term in the sums by, e.g. $\e^{-|x_m|\eps}$.
Since the fermionic functions $\hw_{a,i}^F(x)$ are smooth for
all values of $x$,  the sums over $r\in Z +\ha$ are replaced by
integrals from $-\infty$ to $+\infty$.
However, not all of the  bosonic functions
$\hw_{a,i}^B(x)$   are smooth at $x = -{2\ov \k},
-{1\ov \k}, 0, {1\ov \k}, {2\ov \k}$. Therefore, we obtain the
following formula for the bosonic contribution in \corre
$$
{1\ov 2\kappa}\sum_{n\in Z} \w_n^B =
{1\ov 2\kappa}\big(\w_0^B + 2\w_1^B+2\w_2^B\big)
-\ha\int_{-2/\k}^{3/\k}\ dx\ \big[\k\, {\hw}_{-1}^B(x)+
{1\ov \k}{\hw}_{1}^B(x) + ...\big]  $$
 \eqn\boscor{
+ \
\ha\int_{-\infty}^\infty\ dx\ \big[\k\, {\hw}_{-1}^B(x)+
{1\ov \k}{\hw}_{1}^B(x) + ...\big]\ .
}
Here we have taken into account that $\w_{-n}^B = \w_n^B$,\
$\hw_{0}^B=\hw_{0}^F=0$, and used the result of Appendix C to
replace the two sums, $\sum_{-\infty}^{-3}$ and $\sum_3^\infty$,
by the integrals.

As can be shown by a straightforward computation, the functions
$\hw_{-1}^B$ and $\hw_{-1}^F$, and $\hw_{1}^B$ and
$\hw_{1}^F$, are equal to each other and are given by
\eqn\womo{
\hw_{-1}^B(x) =\hw_{-1}^F(x) = 8\sqrt{ x^2 +1}\ ,}
\eqn\wopo{
\hw_{1}^B(x) =\hw_{1}^F(x) = -\frac{2\,\left( -1 +
3\,\sin^2\g_0 + 2\,x^2\,\sin^2\g_0 \right)}   {\sqrt{{\left(
x^2+ 1 \right) }^3}}\ .}
Therefore, the bosonic and fermionic integrals from $-\infty$ to
$+\infty$ cancel each other up to the order $1\ov \k^2$, and the
correction to the energy is given by the first line in \boscor.

To compute the correction we need to know the large $\k$
expansions of the bosonic frequencies at fixed $n$ (here we set
$\kk=1$). They  are  given up to the
order $1\ov \k$ by the formulas (A.4)--(A.8) of  Appendix A.
Using them, we get the following result
\eqn\corref{
E_1 ={1\ov \k^2} e_1(\g_0) \ ,\ }
\eqn\zag{
e_1(\g_0) \equiv \ha [
5\,\sin^2\g_0-9 + {\sqrt{9 - 12\,\sin^2\g_0}}  +
4\,{\sqrt{4 - 3\,\sin^2\g_0}}] \ .}
Taking into account that at large $\k$
\eqn\zad{  {1\ov \,\k^2} =  {\l\ov (J+2J')^2} + ...\ ,   }
we can rewrite \corref\ in the following form
\eqn\correff{
 E_1 ={\l\ov \,(J+2J')^2}e_1(\g_0) + ... \ .}
At large $\k$ we can also express $\sin^2\g_0$ through the
angular momenta $J$ and $J'$
\eqn\singo{
\sin^2\g_0 \approx {2J'\ov J+2J'} \le {3\ov 4}\ .}
At small values of $\g_0$ or $J'\ll J$ we thus get
$e_1(\g_0) \approx 1$, i.e.
\foot{Note  that  the correction thus does not vanish for $J'=0$.
 This may look as  contradicting  to the fact that at  $\g_0=0$
our solution should be
representing  a BPS state -- a
point particle rotating along a big circle of $S^5$.
As already mentioned above,
 to recover the point-like
case one  should actually  set $\kk$  to 0,
while \corref\ was derived assuming $\kk=1$.
 In general, the $\g_0\to 0$ limit is subtle: expansion near a
point-like string is not a limit of expansion near an extended
string. This is clear from a comparison  of the fluctuation
Lagrangians in the two cases (cf. \fluca). A smooth limit is
found by keeping $\g_0$ arbitrary while sending $\kk=0$:  in this
case we are dealing with ``off-diagonal'' plane of rotation of a
point-like string  and both $J$ and $J'$ are non-zero.}
\eqn\correfff{
E_1 \approx {\l\ov \,(J+2J')^2} \ .}

Combining \corref\ with the classical result  for the  energy
\rty, we obtain the following expression (cf. \sedd)
\eqn\corref{ E = J+2J'+ {\l \ov (J+2J')^2} [ J' + e_1 (\g_0)  +
.. ]
\ .}
We conclude  that the term $J+2J'$ is not modified by the one-loop
sigma model correction. As was discussed above,
 $J'\sim \sql $ is  large in
the semiclassical approximation, and, therefore, the one-loop
sigma model correction is subleading at least at the first order
in $\l$. Since the correction admits an asymptotic expansion in
$1\ov \k^{2m}$ with coefficients depending only on $\sin^2\g_0$,
the one-loop sigma model correction is also subleading at any
order in $\l$.

 It is tempting then to conjecture that all higher-loop sigma
model string corrections are also subleading at large $J'$, and,
therefore, in this regime the classical formula for the energy \rty\  is
exact to all orders in  $\l$.
It should  then be true also at  weak coupling and 
thus should  represent a prediction for
the corresponding anomalous dimensions on the SYM side.
It should be possible to check it using the methods of
\rf{\bis,\mz,\mzn}.


\bigskip\bigskip
\noindent
{\bf Acknowledgements}

\noindent
We are grateful to
G. Arutyunov, R. Metsaev
and K. Zarembo for
useful discussions  and comments.
We thank I. Park for pointing out few  misprints 
in an earlier version of this paper. 
This work  was supported by the DOE grant DE-FG02-91ER40690.
The work of A.T.  was also supported in part by the  PPARC SPG
00613 and  INTAS  99-1590 grants and the Royal Society  Wolfson
award.


\appendix{A}
{Bosonic  frequencies }

Bosonic spectrum of physical fluctuations is determined by zeros
of the determinant \frt\ of the characteristic matrix.
There are also 4 bosons  with masses $m^2 = \k^2$ coming \sqw\ from
the $AdS_5$ part of the background. In what follows $\kk$ will be set
to 1. The dependence on $\kk$ can be restored by rescaling
 $n \to
n/\kk ,\ \ \nu \to \nu/\kk,\ \ \om_n \to \om_n/\kk $.

\subsec{  Expansion at large $n$ }

\noindent To check the ultra-violet finiteness of the model we
need to know the large $n$ expansion of the frequencies up to
the order $1/n$. There are 4 different non-negative frequencies
corresponding to 4 choices of the signs of the two  square roots
in the frequency below
\eqn\ombon{
\w_n = |n| \pm  \sqrt{2+\n^2 \pm \sqrt{(1+\cc)^2+4\n^2\cc}}\ +\
{\n^2\ov 2|n|} + ... \ ,\ \ \ \ \ \ \ \ |n|\gg \nu\ ,}
and 4 frequencies of the  $AdS_5$ fluctuations ($\k^2= \nu^2 + 2
\sin^2 \g_0$) \eqn\omboni{
\w_n^{AdS} = |n| +{\k^2\ov 2|n|} + ... \
,\ \ \ \ \ \ \ \ |n|\gg \nu\ .}
Summing up the contributions of these  4+4=8 frequencies, we get
\eqn\bosnn{
\sum_{i=1}^8 \w_{n,i}^B = 8|n| +4\, {\n^2 + \ss\ov
|n|} + ...  \ .}

\subsec{  Expansion at large $\k$ and fixed $n$}

\noindent
To compute the one-loop correction to the vacuum energy
we need the expansion of the frequencies at large $\k$ (or,
equivalently,  large $\n$)
and fixed $n$. It is given,  up to the order $1/\k$,  by the
following expressions
\eqn\bosomi{
\w_{n,1}^B, \w_{n,2}^B
={1\ov \k}{|n|\ov 2}\,{\sqrt{4 + n^2 - 6\,\sin^2\g_0 \pm
        2\,{\sqrt{4\,n^2\,\cc - 8\,\sin^2\g_0 + 9\,{\sin^4\g_0}}
           }}}    \ ,}
\eqn\bosomiii{
\w_{n,3}^B,\w_{n,4}^B =2\,\kappa + {1\ov 2\,\k}\big(2 + n^2 -
5\,\sin^2\g_0 \pm  {\sqrt{4\,n^2\,\cc  +
         {\sin^4\g_0}}}\, \big)\ ,}
\eqn\bosomv{
\w_{n,i}^B=\kappa + {n^2\ov 2\k}\ ,\ \ \ \ \ \ \ \ \  i=5,6,7,8\
.}

\subsec{  Expansion at large $\k$ and fixed $n\ov \k $}
\noindent
The one-loop computation also requires
the knowledge of the
expansion at large $\k$ and fixed $n\ov \k$. Introducing $x ={
n\ov \k}$, and keeping it fixed, we obtain the following
expansion for $\k \gg 1$ \eqn\bosomki{\w_{n,1}^B, \w_{n,2}^B
=\k ( 1+\sqrt{x^2+1}) \pm  {|x|
\cos\g_0\ov\sqrt{x^2+1}}
+\  {1\ov\k} \big[ \ha\cos 2 \g_0\  -
{2x^2\ss + 3\ss -1\ov 2\sqrt{(x^2+1)^3}}\big] \ ,}
\eqn\bosomkiii{
\w_{n,3}^B, \w_{n,4}^B=\k ( -1+\sqrt{x^2+1}) \pm  {|x|
\cos\g_0\ov\sqrt{x^2+1}}
+\  {1\ov\k}\big[ -\ha\cos 2 \g_0\  -
{2x^2\ss + 3\ss -1\ov 2\sqrt{(x^2+1)^3}}\big] \ ,}
\eqn\bosomkv{
\w_{n,i}^B=
\k \sqrt{x^2+1}\ ,\ \ \ \ \ \ \ \ \ \ \
i=5,6,7,8\ .} Summing up all the 8 frequencies, we get
\eqn\bossumk{
\sum_{i=1}^8\w_{n,i}^B =8\k\sqrt{x^2+1}-
{1\ov\k}\, {2(2x^2\ss + 3\ss -1)\ov \sqrt{(x^2+1)^3}} + ... \ .}

\appendix{B}
{Fermionic  frequencies }

Fermionic spectrum of physical fluctuations is determined (for
$\kk =1$) by zeros of the determinant \zdf\ of the characteristic
matrix. The dependence on $\kk$ can be  restored  again  by rescaling $r
\to r/\kk , \ \nu \to \nu/\kk, \ \om_r \to \om_r/\kk $.

\subsec{  Expansion at large $r$}

\noindent To check that the UV divergences coming from the
bosonic sector (cf. \bosnn) and cancelled by the fermions we
need to know the large $r$ expansion of the frequencies up to
the order $1/r$.
Among the 8 fermionic frequencies there are only 4
different corresponding to 4 choices of the signs of
the 2 square roots in the expression
($r=\pm \ha, ...$)
\eqn\omfermn{
\w_r = |r| \pm \sqrt{1+\n^2+\cc
\pm \n\sqrt{\n^2+2\ss}}\ +\ {\n^2+\ss\ov 2|r|} + ... \ ,\ \ \ \
\ \ \ \ |r|\gg \nu\ .}
Summing up these  8 frequencies, we get
\eqn\fermnn{
\sum_{i=1}^8 \w_{r,i}^F = 8|r| +4\, {\n^2 + \ss\ov
|r|} + ...  \ .}
Comparing this expression with the sum of the bosonic
contributions  \bosnn, we find that indeed the 2-d  UV cancel
 in the 1-loop correction to the energy
\corre.

\bigskip
\subsec{  Expansion at large $\k$ and fixed  $r$}

\noindent
 Even though we do not need the expansion of the
fermionic frequencies at large $\k$ and fixed $r$ to
compute the one-loop correction to the vacuum energy,
 we present
here this expansion for completeness
\eqn\fermmi{
\w^F_{r,1}  ={1\ov 2\k}|r^2-\cc | \ ,}
\eqn\fermmii{
\w_{r,2}^F, \w_{r,3}^F= \kappa  + {1\ov 2\k}\left( r^2 +\cos 2
\g_0\ \pm \sqrt{4\,n^2\,\cc  + \sin^4\g_0}\right) \ ,}
\eqn\fermmiv{
\w_{r,4}^F=2\,\kappa + {1\ov 2\,\k}\left(1 + r^2 - 3\,\sin^2\g_0
\right) \ .}

\bigskip
\subsec{  Expansion at large $\k$ and fixed  $r\ov \k $}

\noindent
 The one-loop computation requires knowledge of the
expansion of the fermionic frequencies
at large $\k$ and
fixed $r/\k$. Introducing $x = r/\k$, and keeping it fixed, we
obtain the following expansion
\eqn\fermmki{
\w_{r,1}^F, \w_{r,2}^F=\k\sqrt{x^2+1} \pm  {|x|
\cos\g_0\ov\sqrt{x^2+1}}- {1\ov\k}\, {x^2\ss +
2\ss -1\ov 2\sqrt{(x^2+1)^3}} \ ,}
\eqn\fermkiii{
\w_{r,3}^F, \w_{r,4}^F
=\k\left( \pm 1+\sqrt{x^2+1}\right) + {1\ov\k}\left(
\pm \ha\cos 2 \g_0\  - {\ss\ov 2\sqrt{x^2+1}}\right) \
.}
Summing up all the eight fermionic  frequencies, we get
\eqn\fersumk{
\sum_{i=1}^8\w_{r,i}^F =8\k\sqrt{x^2+1}-
{1\ov\k}\, {2(2x^2\ss + 3\ss -1)\ov \sqrt{(x^2+1)^3}}+ ... \ .}
Remarkably,
this  expression coincides with the sum of the  bosonic
frequencies  \bossumk.

\appendix{C}
{Approximation of an infinite sum by an integral }

Let us  recall how  sums of the form \corre\ can be
replaced by integrals. Assume that we are given a smooth function
$f(x)$ and its values at points $x_i,\ i=1,\cdots ,N; \ x_{i+1}
- x_i = \D$. We are to find a function $g(x)$ such that the
following formula is valid
\eqn\formu{
\sum_{i=1}^N\ f(x_i) = {1\ov \D} \int_{x_1}^{x_{N+1}}\ dx\ g(x) +
O(\D^5) = {1\ov \D}\sum_{i=1}^N\int_{x_i}^{x_{i+1}}\ dx\ g(x)+
O(\D^5)\ .}
We see that $g(x)$ should satisfy
$$f(x_i) =
{1\ov \D}
\int_{x_i}^{x_{i+1}}\ dx\ g(x) =
g_i + {\D\ov 2}g_i' + {\D^2\ov 6}g_i''+ {\D^3\ov 24}g_i^{(3)} +
{\D^4\ov 120}g_i^{(4)}+ O(\D^5)\ ,$$
where $g_i = g(x_i),\  g_i' = {d\ov dx}g(x_i),\ g_i^{(k)}=
{d^k\ov dx^k}g(x_i)$, and so on.
It is not difficult to check that this formula will be fulfilled
if we choose $g(x)$ to be
\eqn\gx{
g(x) = -{2\ov 15}f(x -\D)+{6\ov 5}f(x -{\D\ov 2})+{1\ov 30}f(x)
-{2\ov 15}f(x +{\D\ov 2})+{1\ov 30}f(x +\D) \ .}
If we are interested in computing the sum in \formu\ only up to
the order $\D^2$ then a simpler formula can be used
\eqn\gxi{
g(x) = {1\ov 3}f(x -\D)+{5\ov 6}f(x)
-{1\ov 6}f(x +\D) \ .}
Note that to use these expressions,  the function $f(x)$ has to be
smooth in the interval $[x_1,x_N]$.


\vfill\eject
\listrefs

\end